\documentclass{article}
\usepackage{amsmath,graphicx,mlspconf,bm}

\toappear{2015 IEEE International Workshop on Machine Learning for Signal Processing, Sept.\ 17--20, 2015, Boston, USA}

\def\Y{{\boldsymbol Y}}
\def\Z{{\boldsymbol Z}}
\def\H{{\boldsymbol H}}
\def\W{{\boldsymbol W}}
\def\V{{\boldsymbol V}}
\def\U{{\boldsymbol U}}
\def\c{{\boldsymbol c}}
\def\x{{\boldsymbol x}}
\def\w{{\boldsymbol w}}
\def\v{{\boldsymbol v}}
\def\u{{\boldsymbol u}}
\def\y{{\boldsymbol y}}
\def\z{{\boldsymbol z}}
\def\h{{\boldsymbol h}}

\def\vSig{{\boldsymbol \Sigma}}
\def\vze{{\boldsymbol \zeta}}
\def\val{{\boldsymbol \alpha}}
\def\S{{\boldsymbol S}}
\def\Q{{\boldsymbol Q}}
\def\tQ{\tilde{\boldsymbol Q}}
\def\vbe{{\boldsymbol \beta}}
\def\vxi{{\boldsymbol \xi}}
\def\vPsi{{\boldsymbol \Psi}}
\def\R{{\boldsymbol R}}
\def\C{{\boldsymbol C}}
\def\tC{\tilde{\boldsymbol C}}
\def\a{{\boldsymbol a}}
\def\B{{\boldsymbol B}}
\def\veta{{\boldsymbol \eta}}
\def\vga{{\boldsymbol \gamma}}
\def\A{{\boldsymbol A}}
\def\b{{\boldsymbol b}}
\def\p{{\boldsymbol p}}
\def\vLam{{\boldsymbol \Lambda}}
\def\I{{\boldsymbol I}}
\def\1{{\boldsymbol 1}}
\def\0{{\boldsymbol 0}}
\def\th{\tilde{\boldsymbol h}}
\def\vphi{{\boldsymbol \phi}}
\def\vPhi{{\boldsymbol \Phi}}
\def\F{{\boldsymbol F}}
\def\vdel{{\boldsymbol \delta}}
\def\vpsi{{\boldsymbol \psi}}

\newcommand{\bc}{\begin{center}}
\newcommand{\ec}{\end{center}}
\newcommand{\be}{\begin{equation}}
\newcommand{\ee}{\end{equation}}
\newcommand{\ba}{\begin{array}}
\newcommand{\ea}{\end{array}}
\newcommand{\bea}{\begin{eqnarray}}
\newcommand{\eea}{\end{eqnarray}}
\newcommand{\bal}{\begin{align}}
\newcommand{\eal}{\end{align}}
\newcommand{\ei}{\end{itemize}}
\newcommand{\bi}{\begin{itemize}}
\newcommand{\bfi}{\begin{figure}}
\newcommand{\efi}{\end{figure}}
\newcommand{\MB}{\left[\begin{array}}
\newcommand{\ME}{\end{array}\right]}
\newcommand{\nn}{\nonumber}

\newcommand{\Exp}{\mathsf{E}}

\newcommand{\Pro}{\mathsf{P}}

\newcommand{\Tr}{\mathsf{Tr}}
\newcommand{\cN}{\mathcal{N}}

\newcommand{\ignore}[1]{{}}

\title{MULTIMODAL FACTOR ANALYSIS}
\name{Yasin Y{\i}lmaz and Alfred O. Hero \thanks{This work was supported in part by the Consortium for Verification Technology under Department of Energy National Nuclear Security Administration award number DE-NA0002534, and the Army Research Office (ARO) grant number W911NF-11-1-0391.}}
\address{University of Michigan, Ann Arbor}
\begin{document}
%\ninept
%

\maketitle
\begin{abstract}
A multimodal system with Poisson, Gaussian, and multinomial observations is considered. A generative graphical model that combines multiple modalities through common factor loadings is proposed. In this model, latent factors are like summary objects that has latent factor scores in each modality, and the observed objects are represented in terms of such summary objects. This potentially brings about a significant dimensionality reduction. It also naturally enables a powerful means of clustering based on a diverse set of observations. An expectation-maximization (EM) algorithm to find the model parameters is provided. The algorithm is tested on a Twitter dataset which consists of the counts and geographical coordinates of hashtag occurrences, together with the bag of words for each hashtag. The resultant factors successfully localizes the hashtags in all dimensions: counts, coordinates, topics. The algorithm is also extended to accommodate von Mises-Fisher distribution, which is used to model the spherical coordinates. 
\end{abstract}
\section{Introduction}
\label{sec:intro}

In complex systems a variety of observation modes (e.g., sensor readings, images, text) may be available to the decision maker. For instance, the emerging technologies, such as cyber-physical systems, internet of things, autonomous driving, and smart grid, can provide such a rich observation space, both in volume and modality. In monitoring a complex system, all modalities bear information about the system's internal state. In some cases, an anomaly may not be detected in each modality alone, but can be easily detected through a joint processing. 

Unsupervised learning methods are instrumental in discovering hidden structures in data (e.g., factor analysis \cite{Zhou12}, topic modeling \cite{Blei03}), which can then be used to perform dimensionality reduction, anomaly detection, and clustering. Conventionally, they deal with unimodal data \cite{Zhou12,Blei03}. Efficient fusion of multimodal data brings about information diversity and can greatly improve the statistical inference performance \cite{Ngiam11}. For example, data used for detection and estimation tasks are strongly coupled when both problems are solved jointly, considerably increasing the overall performance \cite{Yilmaz14}.

 In \cite{Khan10}, Gaussian and multinomial observations are jointly modeled using a mixture of factor analyzers. Although that work is conceptually similar to our work, its probabilistic model is quite different. In \cite{Khan10}, the latent factor scores are shared in the system, which also synchronizes the modalities. Here we do not impose a synchronized observation model. We let the different modalities run in their own continuous-valued state, and only link them through the factor loadings of objects. On the contrary, in \cite{Khan10}, modalities differ in their factor loadings, which can take a discrete set of values.

\section{Problem Statement}
\label{sec:problem}

We consider $P$ objects (e.g., documents) each of which is observed through three disparate information sources. In this paper, we assume Poisson, Gaussian, and multinomial statistical models for the information sources because of their wide application areas. Specifically, Poisson distribution is used to model event counts; Gaussian distribution is used for continuous-valued observations; and multinomial distribution models categorical observations. (Real-world examples can be seen in the Experiments section.) 

\bfi
\centering
\includegraphics[width=3.5in]{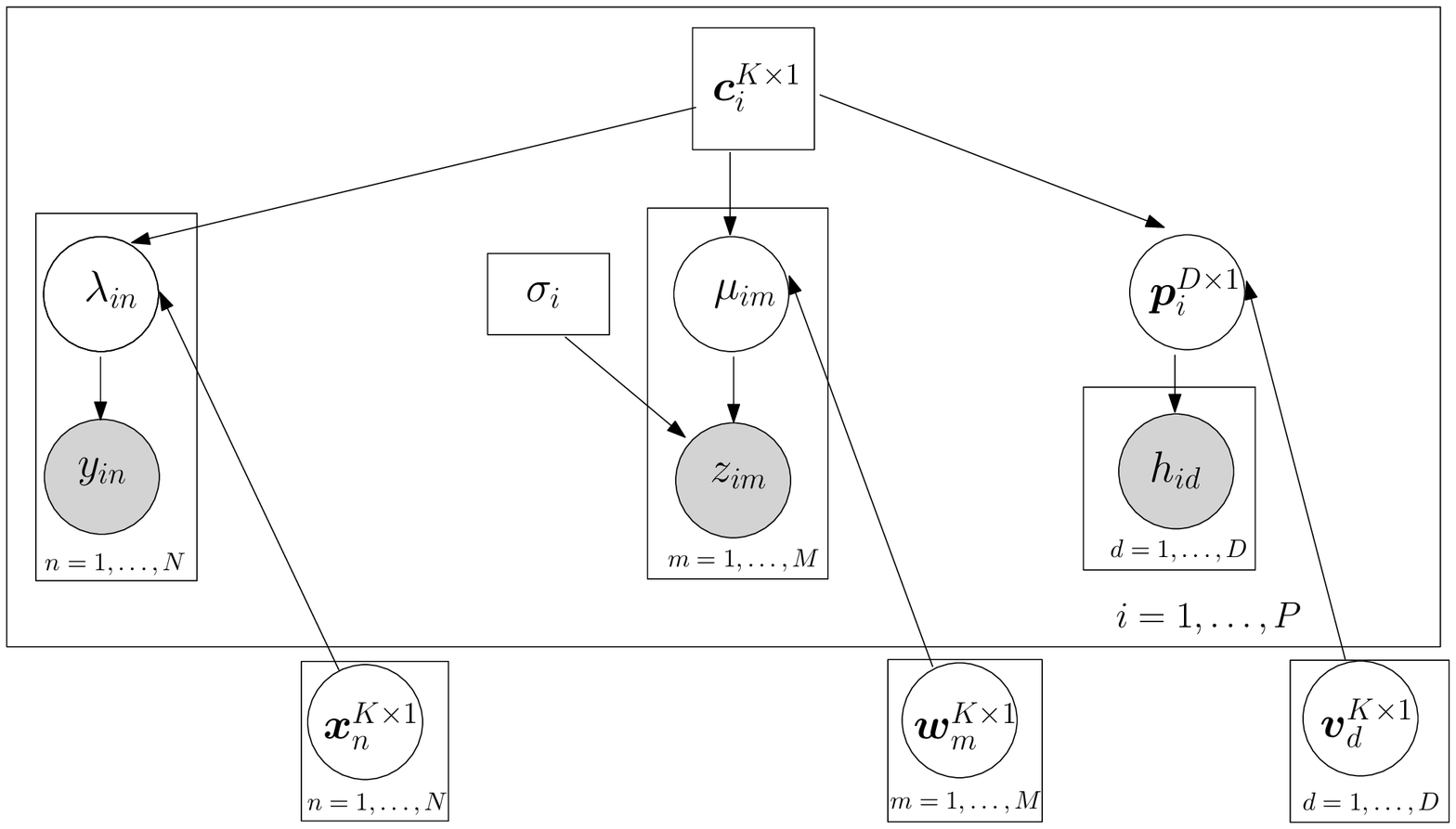}
\caption{Generative graphical model. Plate representation is used to show repeated structures. Circles  and rectangles represent random and deterministic variables, respectively. Observed variables are shaded. Each object $i$ has three disparate observation streams: Poisson $\{y_{in}\}_n$, Gaussian $\{z_{im}\}_m$, and multinomial $\{h_{id}\}_d$. Factor loadings $\c_i$, and the latent factor scores $\x_n$, $\w_m$, and $\v_d$ constitute the multimodal factor model.}
\label{fig:model}
\efi

The graphical model in Fig. \ref{fig:model} is assumed to generate the multimodal observations $\Y^{P \times N}$, $\Z^{P \times M}$, and $\H^{P \times D}$. For each object $i$, the latent factor scores $\x_n$, $\w_m$, and $\v_d$ for $K$ factors are mixed in the the natural parameters of the distributions through the unknown factor loadings $\c_i$. Gaussian priors are assumed for the latent factor scores. In particular, each Poisson observation $y_{in}$ is conditionally distributed as
\begin{align}
\label{eq:pois}
	y_{in} | \x_n &\sim \text{Pois}(e^{\c_i^T\x_n}), \\
	\x_n &\sim \cN(\vze,\R) \nn	
\end{align}
for $i = 1,\ldots,P, ~ n = 1,\ldots,N$. Similarly, for each Gaussian observation $z_{in}$ we have
\begin{align}
\label{eq:gaus}
	z_{im} | \w_m &\sim \cN(\c_i^T\w_m,\sigma_i^2), \\
	\w_m &\sim \cN(\val,\S) \nn
\end{align}
for $m = 1,\ldots,M$. We use different observation indices for different information sources since they do not need to be synchronized and the total number of observations may vary. 
For each multinomial observation $h_{id}$ we have
\begin{align}
\label{eq:mult}
	h_{id} | \{\v_d\} &\sim \text{Mult}\left(L_i; \frac{e^{\c_i^T \v_1}}{\sum_{d=1}^D e^{\c_i^T \v_d}}, \ldots, \frac{e^{\c_i^T \v_D}}{\sum_{d=1}^D e^{\c_i^T \v_d}} \right), \\
	\v_d &\sim \cN(\vbe,\Q), \nn
\end{align}
where $L_i$ is the total number of instances and $h_{id}$ is the number of instances observed under category $d$. 

The above probabilistic models are similar to generalized linear models since mixing occurs in the natural parameters. However, here the regressors $\x_n$, $\w_m$, and $\v_d$ are unknown, as well as the regression coefficients $\c_i$, as opposed to generalized linear models.

\section{Expectation-Maximization Algorithm}
\label{sec:em}

In this section, we derive the expectation-maximization (EM) algorithm to find the parameters 
\be
	\theta = \{\c_i,\vze,\R,\sigma_i^2,\val,\S,\vbe,\Q\}. \nn
\ee

\subsection{Poisson E-step}
\label{sec:poisE}

We are interested in computing the expectation of the complete-data log-likelihood $\Exp\left[ \log \Pro(\{\y_n,\x_n\} | \theta) \right]$ over the posterior distribution $\Pro(\{\x_n\} | \{\y_n\},\theta)$ of the latent factor scores. Due the lack of conjugacy between the prior on $\x_n$ and the Poisson likelihood, the posterior distribution does not have a closed form expression. Therefore, as in \cite{Smith03}, we approximate the posterior with a Gaussian whose mean and covariance are the mode of the posterior (i.e., MAP estimate of $\x_n$) and the negative inverse Hessian of the log posterior at that mode. 

We start by writing
\begin{multline}
	\log \Pro(\x_n|\y_n,\theta) = \log \Pro(\y_n|\x_n,\theta) + \log \Pro(\x_n|\theta) + C_1 \\
	= \left( \sum_{i=1}^P -e^{\c_i^T\x_n} + y_{in}\c_i^T\x_n \right) \\ - \frac{1}{2} \x_n^T \R^{-1} \x_n + \vze^T \R^{-1} \x_n + C_2 \nn,
\end{multline}
where $C_1$ and $C_2$ are constants that do not depend on $\x_n$, and we used the Poisson likelihood from \eqref{eq:pois} and the multivariate Gaussian pdf. Then, the gradient and the Hessian are given by
\begin{multline}
	\nabla_{\x_n} \log \Pro(\x_n|\y_n,\theta) = \\ \sum_{i=1}^P \left(-e^{\c_i^T\x_n} + y_{in} \right) \c_i - \R^{-1} (\x_n-\vze)  \\
	\nabla^2_{\x_n} \log \Pro(\x_n|\y_n,\theta) = \sum_{i=1}^P -e^{\c_i^T\x_n} \c_i\c_i^T - \R^{-1}. \label{eq:hess}
\end{multline}
Since $\log \Pro(\x_n|\y_n,\theta)$ is strictly concave in $\x_n$, we can find the unique mode $\vxi_n$ using the gradient and the Hessian in Newton's method, i.e.,
\begin{align}
	\mathcal{H}_{\x^{t}_n} (\x^{t+1}_n-\x^{t}_n) &= -\mathcal{G}_{\x^{t}_n}, \nn ~~~~
	\x^t_n \to \vxi_n, \nn
\end{align}
where $\mathcal{G}_{\x^{t}_n}$ and $\mathcal{H}_{\x^{t}_n}$ are the gradient and the Hessian at iteration $t$, computed using \eqref{eq:hess}.

We approximate the posterior as \newline $\Pro(\x_n|\y_n,\theta) \approx \cN(\vxi_n,\vPsi_n)$, where $\vPsi_n = -\mathcal{H}_{\vxi_n}^{-1}$. Finally, the expected complete-data log-likelihood is given by
\begin{multline}
	\Exp\left[ \log \Pro(\{\y_n,\x_n\} | \theta) \right] \\
	= \sum_{n=1}^N \Exp\left[ \log \Pro(\y_n|\x_n,\theta) + \log \Pro(\x_n|\theta) \right] \\
	= \sum_{n=1}^N \sum_{i=1}^P \Exp\left[-e^{\c_i^T\x_n} + y_{in}\c_i^T\x_n \right] - \frac{1}{2} \Exp\left[ \x_n^T \R^{-1} \x_n \right] \\
	+ \vze^T \R^{-1} \Exp\left[ \x_n \right] - \frac{1}{2} \vze^T \R^{-1} \vze - \frac{1}{2} |\R| + C_3 \\
	\approx \sum_{n=1}^N \sum_{i=1}^P \left(-\Exp\left[e^{\c_i^T\x_n}\right] + y_{in}\c_i^T\vxi_n \right) \\
	- \frac{1}{2} \Tr\left( \R^{-1} (\vPsi_n+\vxi_n\vxi_n^T) \right) + \vze^T \R^{-1} \vxi_n \\
	- \frac{1}{2} \vze^T \R^{-1} \vze - \frac{1}{2} |\R| + C_3,
	\label{eq:pois_max}
\end{multline}
where $C_3$ is a constant and we used the fact that 
\begin{align}
	\Exp\left[ \x_n^T \R^{-1} \x_n \right] &= \Exp\left[ \Tr(\x_n^T \R^{-1} \x_n) \right] = \Exp\left[ \Tr(\R^{-1} \x_n\x_n^T) \right] \nn\\
	&=  \Tr\left( \R^{-1} \Exp\left[\x_n\x_n^T \right] \right) \nn\\
	&\approx \Tr\left( \R^{-1} (\vPsi_n+\vxi_n\vxi_n^T) \right). 
	\label{eq:quad}
\end{align}

\subsection{Poisson M-step}
\label{sec:poisM}

In iteration $t+1$, we find the parameters $\vze^{t+1}$ and $\R^{t+1}$ that maximize \eqref{eq:pois_max}. Particularly,
\be
	\vze^{t+1} = \arg\max_{\vze} \vze^T \R^{-1} \sum_{n=1}^N \vxi^t_n - \frac{N}{2} \vze^T \R^{-1} \vze. \nn
\ee
Equating the derivative to zero we find that the mean of the factor scores is given by the average of the posterior means, i.e., 
\be
	\vze^{t+1} = \frac{1}{N} \sum_{n=1}^N \vxi^t_n.
\ee

Similarly,
\begin{multline}
	\R^{t+1} = \arg\max_{\vze} \sum_{n=1}^N -\frac{1}{2} \Tr\left( \R^{-1} (\vPsi^t_n+\vxi^t_n(\vxi^t_n)^T) \right) \\+ \vze^T \R^{-1} \sum_{n=1}^N \vxi^t_n - \frac{N}{2} \vze^T \R^{-1} \vze - \frac{N}{2} |\R|. \nn
\end{multline}
Taking the derivative and equating to zero we get
\begin{multline}	
	\sum_{n=1}^N (\R^{t+1})^{-1} \Bigg[ \frac{1}{2} (\vPsi^t_n+\vxi^t_n(\vxi^t_n)^T) - \vze^{t+1}\vxi_n^T \\
	+ \frac{1}{2}\vze^{t+1}(\vze^{t+1})^T \Bigg] (\R^{t+1})^{-1} - \frac{N}{2} (\R^{t+1})^{-1} = 0 \\
	\R^{t+1} = \frac{1}{N} \sum_{n=1}^N \left( \vPsi^t_n+\vxi^t_n(\vxi^t_n)^T \right) - \vze^{t+1}(\vze^{t+1})^T.
	\label{eq:pois_cov}
\end{multline}

\subsection{Gaussian E-step}
\label{sec:gausE}

We want to compute $\Exp\left[ \log \Pro(\{\z_m,\w_m\} | \theta) \right]$ over the posterior distribution $\Pro(\{\z_m\} | \{\w_m\},\theta)$ of the latent factor scores. Since the Gaussian prior on the factor scores is conjugate to the Gaussian likelihood, the posterior is also Gaussian. To find its mean and covariance, from \eqref{eq:gaus}, we write
\begin{multline}
	\Pro(\z_m,\w_m|\theta) = \Pro(\z_m|\w_m,\theta) \Pro(\w_m|\theta) \\
	= \frac{e^{ -\frac{1}{2} \left[ (\z_m-\C\w_m)^T \vSig^{-1} (\z_m-\C\w_m) + (\w_m-\val)^T \S^{-1} (\w_m-\val) \right]} }{(2\pi)^{\frac{P+K}{2}} |\vSig|^{\frac{1}{2}} |\S|^{\frac{1}{2}} }, \nn
\end{multline}
where $\C=[\c_1 \ldots \c_P]^T$ and $\vSig=\text{diag}(\sigma^2_1,\ldots,\sigma^2_P)$. Collecting the terms that depend on $\z_m$ together and completing the square we finally obtain the posterior mean and covariance as
\be
	\a_m = \B (\C^T \vSig^{-1} \z_m + \S^{-1} \val), ~~\B = (\C^T \vSig^{-1} \C + \S^{-1})^{-1},
	\label{eq:gaus_post}
\ee
respectively. 

Then, the expected complete-data log-likelihood is written as
\begin{multline}
	\Exp\left[ \log \Pro(\{\z_m,\w_m\} | \theta) \right] \\
	= \sum_{m=1}^M \Exp\left[ \log \Pro(\z_m|\w_m,\theta) + \log \Pro(\w_m|\theta) \right] \\
	= -\frac{1}{2} \Bigg\{ \sum_{m=1}^M \sum_{i=1}^P \left( \frac{1}{\sigma^2_i} \Exp\left[ (z_{im}-\c_i^T\w_m)^2 \right] + \log(2\pi\sigma^2_i) \right) \\
	+ \Exp\left[ (\w_m-\val)^T \S^{-1} (\w_m-\val) \right] + \log((2\pi)^K|\S|) \Bigg\}.
	\label{eq:gaus_max}
%:
\end{multline}

\subsection{Gaussian M-step}
\label{sec:gausM}

At each iteration $t+1$, we find the parameters $\val^{t+1}$, $\S^{t+1}$, and $\{(\sigma^2_i)^{t+1}\}_i$ that maximize \eqref{eq:gaus_max}. Specifically,
\begin{align}
	\val^{t+1} &= \arg\max_{\val} -\frac{1}{2}\sum_{m=1}^M \Exp\left[ (\w_m-\val)^T \S^{-1} (\w_m-\val) \right] \nn\\
	&= \arg\max_{\val} \val^T \S^{-1} \sum_{m=1}^M \a^t_m -\frac{M}{2} \val^T \S^{-1} \val, \nn
\end{align}
where $\a^t_m$, given in \eqref{eq:gaus_post}, is the posterior mean $\Exp[\w_m]$ at iteration $t$. Taking the derivative we get
\be
	\val^{t+1} = \frac{1}{M} \sum_{m=1}^M \a^t_m.
\ee

Similarly, 
\begin{multline}
	\S^{t+1} = \arg\max_{\S} \\
	-\frac{1}{2} \sum_{m=1}^M \Exp\left[ (\w_m-\val)^T \S^{-1} (\w_m-\val) \right] -\frac{M}{2} \log|\S| \\
	= \arg\max_{\val} -\frac{1}{2} \sum_{m=1}^M \Tr\left( \S^{-1} (\B+\a_m\a_m^T) \right) \\
	+ \val^T \S^{-1} \sum_{m=1}^M \a^t_m -\frac{M}{2} \val^T \S^{-1} \val -\frac{M}{2} \log|\S|, \nn
\end{multline}
where we used \eqref{eq:quad} to write $\Exp\left[ \w_m^T \S^{-1} \w_m \right]$. Similar to \eqref{eq:pois_cov}, taking the derivative we find 
\be
	\S^{t+1} = \frac{1}{M} \sum_{m=1}^M \left( \B^t_m+\a^t_m(\a^t_m)^T \right) - \val^{t+1}(\val^{t+1})^T.
	\label{eq:gaus_cov}
\ee

We next find
\begin{multline}
	(\sigma^2_i)^{t+1} = \arg\max_{\sigma^2_i} -\frac{1}{2\sigma^2_i} \sum_{m=1}^M \Exp\left[ (z_{im}-\c_i^T\w_m)^2 \right] \\
	-\frac{M}{2\sigma^2_i} \log \sigma^2_i, \nn
\end{multline}
where, from \eqref{eq:gaus_post}, $\c_i^T\w_m|\z_m \sim \cN(\c_i^T\a_m,\c_i^T\B\c_i)$, hence $\Exp\left[ (z_{im}-\c_i^T\w_m)^2 \right] = \c_i^T\B\c_i+(\c_i^T\a_m-z_{im})^2$. Equating the derivative to zero and solving for $\sigma^2_i$ we get
\be
	(\sigma^2_i)^{t+1} = \frac{1}{M} \sum_{m=1}^M (z_{im}-\c_i^T\a_m)^2 + \c_i^T\B\c_i.
\ee

\subsection{Multinomial E-step}
\label{sec:multE}

In the multinomial case, for identifiability, we use the last category as pivot and write the likelihood in terms of the alternative factor scores $\u_d = \v_d-\v_D,~ d=1,\ldots,D$,
\begin{align}
	\Pro(\h_i|\{\u_d\},\theta) &= \prod_{d=1}^{D} \left(\frac{e^{\c_i^T\u_d}}{1+\sum_{l=1}^{D-1} e^{\c_i^T\u_l}}\right)^{h_{id}} \nn\\
	&= \prod_{d=1}^{D} e^{[\eta_{id}-\text{lse}(\veta_i)]h_{id}}, 
	\label{eq:mult_like}
\end{align}
where $\eta_{id} = \c_i^T\u_d$ and $\veta_i=[\eta_{i1} \ldots \eta_{iD-1}]$. The normalizing term in the probability expression, also called the log-sum-exp (lse) function prevents a closed form solution for the posterior. Finding a quadratic upper bound for it we can bound from below the likelihood, and in turn the expected complete-data log-likelihood, which we want to maximize. 

Using the Taylor series of lse$(\veta_i)$ we can find such a quadratic bound \cite{Bohning92} as follows,
\begin{multline}
	\text{lse}(\veta_i) = \text{lse}(\vga_i) + (\veta_i-\vga_i)^T \nabla \text{lse}(\vga_i) \\
	+ \frac{1}{2} (\veta_i-\vga_i)^T \nabla^2 \text{lse}(\vga_i+\epsilon(\veta_i-\vga_i)) \\
	\le \frac{1}{2} \veta_i^T \A \veta_i - \b_{\vga_i}^T \veta_i + c_{\vga_i}.
	\label{eq:lse}
\end{multline}
To show the above inequality note that $\nabla \text{lse}(\vga_i)$ is the probability vector $\p_i(\vga_i)$, and $\nabla^2 \text{lse} = \vLam_{\p_i}-\p_i\p_i^T$ where $\vLam_{\p_i}=\text{diag}(p_{i1},\ldots,p_{iD-1})$. In \cite{Bohning92}, the latter is shown to be bounded as follows
\be
	\nabla^2 \text{lse} \le \A = \frac{1}{2}\left( \I_{D-1} - \frac{\1_{D-1}\1_{D-1}^T}{D} \right), \nn
\ee
where $\I_{D-1}$ and $\1_{D-1}$ are the identity matrix and the vector of ones of size $D-1 \times D-1$ and $D-1 \times 1$, respectively. Substituting $\A$ and $\nabla \text{lse}(\vga_i)$ and organizing the terms gives us the inequality in \eqref{eq:lse}, where
\be
	\b_{\vga_i} = \A\vga_i - \p_i(\vga_i), ~~ c_{\vga_i} = \text{lse}(\vga_i) + \frac{1}{2} \vga_i^T \A \vga_i - \p_i(\vga_i). \nn
\ee

From \eqref{eq:mult_like} and \eqref{eq:lse}, we get
\begin{multline}
	\log \Pro(\h_i|\{\u_d\},\theta) \ge \h_i^T \veta_i - \left(\frac{1}{2} \veta_i^T \A \veta_i - \b_{\vga_i}^T \veta_i + c_{\vga_i} \right) L_i \\
	= -\frac{1}{2} \left( \veta_i-\A^{-1}\left( \frac{\h_i}{L_i}+\b_{\vga_i} \right) \right)^T \A \\
	\left( \veta_i-\A^{-1}\left( \frac{\h_i}{L_i}+\b_{\vga_i} \right) \right) \\
	+ \frac{1}{2} \left( \frac{\h_i}{L_i}+\b_{\vga_i} \right)^T \A^{-1} \left( \frac{\h_i}{L_i}+\b_{\vga_i} \right) - c_{\vga_i}. \nn
\end{multline}
Exponentiating we obtain the following lower bound for the likelihood
\be
	\Pro(\h_i|\{\u_d\},\theta) \ge \cN(\th_i|\veta_i,\A^{-1}) f_i(\vga_i), \nn
\ee
where $\th_i = \A^{-1}\left( \frac{\h_i}{L_i}+\b_{\vga_i} \right) = \A^{-1}\left( \frac{\h_i}{L_i}-\p_i(\vga_i) \right) + \vga_i$ is the Gaussian pseudo-observation

Since the factor scores $\{\u_d\}$ are correlated given the observations, we seek the posterior of the combined vector \newline $\u = [u_{11} \ldots u_{D-1~K}]^T \sim \cN(\0,\tQ)$ where $\0$ is the zero vector and $\tQ = \I_{D-1} \otimes Q$ is a block-diagonal matrix. Similarly, defining $\tC_i = \I_{D-1} \otimes \c_i$ we can write $\veta_i=\tC_i^T\u$. Then, for the complete-data likelihood we have
\begin{multline}
	\Pro(\{\h_i\},\u|\theta) \ge \\
	\left( \prod_{i=1}^P \cN(\th_i|\tC_i^T\u,\A^{-1}) f_i(\vga_i) \right) \cN(\u|\0,\tQ).
	\label{eq:mult_low}
\end{multline}
Organizing the terms and completing the square we write
\be
	\Pro(\{\h_i\},\u|\theta) \ge \cN(\u|\vphi,\vPhi) ~g_i(\{\h_i,\vga_i\}), \nn
\ee
where \newline 
\be
	\vphi = \vPhi \sum_{i=1}^P \tC_i\A\th_i, ~~\vPhi = \left(\sum_{i=1}^P \tC_i\A\tC_i^T+\tQ^{-1} \right)^{-1} 
	\label{eq:mult_post}
\ee
are the posterior mean and covariance. 

\subsection{Multinomial M-step}
\label{sec:multM}

We maximize the expected complete-data log-likelihood of the lower bound given in \eqref{eq:mult_low} using the posterior mean and covariance, given in \eqref{eq:mult_post}, as in \cite{Khan10}.
\begin{multline}
\label{eq:mult_max}
	\Q^{t+1} = \arg\max_{\Q} \Exp\left[ \log \Pro(\{\h_i\},\u|\theta) \right] \\
	= \arg\max_{\Q} -\frac{1}{2} \Exp\left[\u^T \tQ^{-1} \u\right] - \frac{D-1}{2} \log|\Q| \\
	-\frac{1}{2} \Exp\left[\u^T \sum_{i=1}^P \tC_i\A\tC_i^T \u\right] + \Exp[\u]^T \sum_{i=1}^P \tC_i\A\th_i \\
	-\frac{1}{2} \sum_{i=1}^P \th_i^T\A\th_i + \sum_{i=1}^P \log f_i(\vga_i) + C_4 \\
\end{multline}
\begin{multline}
	\Q^{t+1} = \arg\max_{\Q} -\frac{1}{2} \Tr\left( \tQ^{-1} (\vPhi+\vphi\vphi^T) \right) \\
	- \frac{D-1}{2} \log|\Q| \\
	= \arg\max_{\Q} -\frac{1}{2} \sum_{d=1}^{D-1} \Tr\left( \Q^{-1} (\vPhi_d+\vphi_d\vphi_d^T) \right) \\
	- \frac{D-1}{2} \log|\Q|
\end{multline}
where $C_4$ is a constant, $\vphi_d$ is the $d$th vector of size $K$ in $\vphi$, and $\vPhi_d$ is the $d$th matrix of size $K \times K$ on the diagonal of $\vPhi$. The last equality follows from the fact that $\tQ$ is block diagonal. We used \eqref{eq:quad} for $\Exp\left[\u \tQ^{-1} \u\right]$. Similar to \eqref{eq:pois_cov} and \eqref{eq:gaus_cov} we find
\be
	\Q^{t+1} = \frac{1}{D-1} \sum_{d=1}^{D-1} \left( \vPhi^t_d+\vphi^t_d(\vphi^t_d)^T \right).
\ee

Since \eqref{eq:lse} holds with equality for $\vga_i=\veta_i$, and the curvature does not depend on $\veta_i$, it can be shown that the optimal value for $\vga_i$ is $\tC_i^T\vphi$ \cite{Khan10}. Note that using the factor scores $\u_d\sim \cN(\0,\Q)$ the mean vector $\vbe$ in \eqref{eq:mult} is not needed. 

\subsection{Factor Loadings}
\label{sec:factor}

Finally, combining \eqref{eq:pois_max}, \eqref{eq:gaus_max} and \eqref{eq:mult_max} we compute $\c_i^{t+1}$ as follows
\begin{multline}
	\c_i^{t+1} = \arg\max_{\c_i} \Exp[\{\y_n,\x_n,\z_m,\w_m,\h_i,\u\}|\theta] \\
	= \arg\max_{\c_i} \sum_{n=1}^N \left(-\Exp\left[e^{\c_i^T\x_n}\right] + y_{in}\c_i^T\vxi_n \right) \\
	-\frac{M}{2\sigma^2_i} \c_i^T\B\c_i -\frac{1}{2\sigma^2_i} \sum_{m=1}^M  (\c_i^T\a_m-z_{im})^2 \\
	-\frac{1}{2} \Tr\left( \tC_i\A\tC_i^T (\vPhi+\vphi\vphi^T)\right) + \vphi^T \tC_i\A\th_i.
\end{multline}
Note that $\c_i^T\x_n | \{\y_n\} \sim \cN(\c_i^T\xi_n,\c_i^T\Psi_n\c_i)$. Completing the square in the Gaussian integral it is straightforward to show that $\Exp\left[e^{\c_i^T\x_n}\right] = e^{\c_i^T\xi_n+\frac{\c_i^T\Psi_n\c_i}{2}}$. 
Moreover, we can write 
$\Tr\left( \tC_i\A\tC_i^T (\vPhi+\vphi\vphi^T)\right) = \c_i^T \U \c_i$, and $\vphi^T \tC_i\A\th_i = \c_i^T \vdel_i$, where $\U = \frac{D-1}{2}\Q^{t+1} - \V$, $\V=[\vpsi_1 \cdots \vpsi_K]$, 
$\vpsi_k = \frac{1}{2D}\sum_{\ell=0}^{D-2} \sum_{d=1}^{D-1} $ and $\F=[\vphi_1 \ldots \vphi_{D-1}]$. 

We can use Newton's method to find $\c_i^{t+1}$ through the gradient and the Hessian, i.e.,
\begin{multline}
	\nabla_{\c_i} = -\sum_{n=1}^N e^{\c_i^T\xi_n+\frac{\c_i^T\Psi_n\c_i}{2}} (\xi_n+\Psi_n\c_i) + \sum_{n=1}^N y_{in}\vxi_n \\
	-\frac{M}{\sigma^2_i} \B\c_i - \frac{1}{\sigma^2_i} \sum_{m=1}^M  (\c_i^T\a_m-z_{im}) \a_m \\
	-\sum_{d=1}^{D-1} a_d (\vPhi_d+\vphi_d\vphi_d^T) \c_i + \F \A \th_i \\	
	\nabla^2_{\c_i} = -\sum_{n=1}^N e^{\c_i^T\xi_n+\frac{\c_i^T\Psi_n\c_i}{2}} [(\xi_n+\Psi_n\c_i)(\xi_n+\Psi_n\c_i)^T + \vPsi_n] \\
	- \frac{1}{\sigma^2_i} \sum_{m=1}^M  \a_m\a_m^T -\sum_{d=1}^{D-1} a_d (\vPhi_d+\vphi_d\vphi_d^T).
\end{multline}

\section{Experiments}
\label{sec:exper}

We test our algorithm on a Twitter dataset that is filtered from the 10\% of the tweets in January 2013. In our dataset, we consider 2444 hashtags as objects (i.e., $P=2444$); and analyze their number of occurrences in 743 hours (i.e., $N=743$), available geographical coordinates, and word counts. We model the hashtag occurrences (i.e., number of tweets that mention a hashtag) using Poisson distribution. Word counts are modeled using multinomial distribution with a dictionary size of 2645 after eliminating the words that appear less than 100 times in the whole dataset.

The number of available coordinates ranges between 10 and 10456 with a mean of 134. Since geographical coordinates (latitude and longitude) constitute spherical data, they are better modeled using von Mises-Fisher (vMF) distribution than Gaussian, which is treated initially due to its popularity. Thus, we here present an extension of our algorithm for vMF distribution. The observation model for vMF distribution is given below
\begin{align}
	\z_{im}^T | \W_m \sim& \text{vMF}(\c_i^T \W_m,\kappa_i), \nn\\
	\w_{mk}^T \sim& \text{vMF}(\val_k,s_k), \nn
\end{align}
for $m=1,\ldots,M_i$, where $\z_{im}^{3\times1}$ is the spherical coordinate vector obtained from the original latitude and longitude information; \newline $\W_m^{K\times3} = [\w_{m1} \ldots \w_{mK}]^T$ is the latent factor scores for all three dimensions. The scores $\w_{mk}^{3\times1}$ for each factor $k$ are also vMF distributed. 

The vMF likelihood is given by
\be
	\Pro(\z_{im}|\W_m,\theta) = C(\kappa_i) e^{\kappa_i \W_m^T \c_i \z_{im}}, \nn
\ee
where $C(\kappa_i) = \frac{\kappa_i}{2\pi(e^\kappa_i-e^{-\kappa_i})}$, $\kappa_i$ is the concentration parameter, and $\c_i^T \W_m$ is the mean direction. Combining the likelihood with the prior it is straightforward to show that the posterior $\Pro(\w_{mk}|\z_{im},\theta)$ is also vMF with mean and concentration given by
\begin{align}
	\a_{mk} = \frac{s_k \val_k + \kappa_i c_{ik} \z_{im}}{b_{mk}}~, ~~ b_{mk} = \|s_k \val_k + \kappa_i c_{ik} \z_{im}\|, \nn
\end{align}
respectively. Maximizing the expected complete-data log-likelihood $\Exp[\log \Pro(\{\z_{im},\w_{mk}\}|\theta)]$ we get
\be
	\val_k^{t+1} = \frac{\sum_{m=1}^{M}\a_{mk}^t}{\|\sum_{m=1}^{M} \a_{mk}^t\|}, ~~ M = \max(M_i). \nn
\ee
Using the well-known approximation we estimate the concentration parameters as
\begin{align}
	s_k^{t+1} &= \frac{3\bar{r}_k-\bar{r}_k^3}{1-\bar{r}_k^2}, ~~ \bar{r}_k = \frac{\|\sum_{m=1}^{M} \a_{mk}^t\|}{M} \nn\\
	\kappa_i &= \frac{3\bar{R}_i-\bar{R}_i^3}{1-\bar{R}_i^2}, ~~ \bar{R}_i = \frac{\|\sum_{m=1}^{M_i} \z_{im}\|}{M_i}. \nn
\end{align}
We choose the number of factors through experiments based on the perplexity performance for the Poisson observations. 

Table \ref{tab:htag} lists the dominant hashtags in five selected factors. 
\begin{table}[b]
\caption{Factors localized in terms of popularity, geography, and topic. Numbers in parentheses denote the average popularity scores.}
\begin{center}
\begin{tabular}{|c|c|c|c|c|}
\hline Europe Soccer & US New Year & Middle East Countries & Astrology & East Coast Sports \\ \hline\hline
\#CFC (83) & \#HappyNewYear (49) & \#bahrain (133) & \#Aries (210) & \#Patriots (29) \\ \hline
\#FACup (40) & \#NewYear (13) & \#Pakistan (30) & \#Pisces (231) & \#Knicks (18) \\ \hline
\#Arsenal (33) & \#newyearseve (3) & \#India (11) & \#Capricorn (210) & \#GoHawks (20) \\ \hline
\#ballondor (17) & \#Best2012Memories (4) & \#kuwait (5) & \#Virgo (192) & \#Steelers (2) \\ \hline
\#realmadrid (14) & \#Feliz2013 (1) & \#Iraq (2) & \#Sagittarius (216) & \#Yankees (3) \\ \hline
\end{tabular}
\end{center}
\label{tab:htag}
\end{table}
Our algorithm localizes factors in three aspects: popularity, geography, and topic. On the other hand, a unimodal factor analyzer handles only one of those. For example, the first factor in Table \ref{tab:htag} is about European soccer, and the dominant hashtags in this factor are all popular. Average number of occurrences over $743$ hours are given in parentheses. Similarly hashtags related to the second factor share the common topic of new year, and are on average tweeted from US. Third factor is about Middle East countries. The reason Bahrain was so popular in January 2013 seems to be the Gulf Cup of Nations, a soccer tournament organized in Bahrain between January 5th and 18th. Astrology hashtags have regularly high counts. The last factor focuses on the East Coast sports teams in football, basketball, and baseball. 

\section{Conclusion}
\label{sec:conc}

A generative graphical model and an EM algorithm to analyze it have been proposed to be able summarize multimodal objects that consist of disparate observations. In a one-month Twitter dataset, the discovered factors, each of which acts as a summary multimodal object, have been shown to successfully localize hashtags in terms of popularity, geography, and topic.

\bibliographystyle{IEEEbib}
\bibliography{refs}

\begin{thebibliography}{1}

\bibitem{Zhou12}
Mingyuan Zhou, Lauren~A. Hannah, David~B. Dunson, and Lawrence Carin,
\newblock ``Beta-negative binomial process and poisson factor analysis,''
\newblock in {\em AISTATS}, 2012.

\bibitem{Blei03}
David~M. Blei, Andrew~Y. Ng, and Michael~I. Jordan,
\newblock ``Latent dirichlet allocation,''
\newblock {\em J. Mach. Learn. Res.}, vol. 3, pp. 993--1022, Mar. 2003.

\bibitem{Ngiam11}
J.~Ngiam, A.~Khosla, B.~Kim, J.~Nam, H.~Lee, and A.Y. Ng,
\newblock ``Multimodal deep learning,''
\newblock in {\em ICML}, 2011.

\bibitem{Yilmaz14}
Y.~Y{\i}lmaz, G.V. Moustakides, and X.~Wang,
\newblock ``Sequential joint detection and estimation,''
\newblock {\em Theory of Probability \& Its Applications}, vol. 59, no. 3,
  2014.

\bibitem{Khan10}
M.E. Khan, G.~Bouchard, B.M. Marlin, and K.P. Murphy,
\newblock ``Variational bounds for mixed-data factor analysis,''
\newblock in {\em Neural Information Processing Systems (NIPS) Conference},
  2010.

\bibitem{Smith03}
A.C. Smith and E.N. Brown,
\newblock ``Estimating a state-space model from point process observations,''
\newblock {\em Neural Comput.}, vol. 15, no. 5, pp. 965--991, 2003.

\bibitem{Bohning92}
D~B\"{o}hning,
\newblock ``Multinomial logistic regression algorithm,''
\newblock {\em Ann. Inst. Statist. Math.}, vol. 44, no. 1, pp. 197--200, 1992.

\end{thebibliography}

\end{document}